\documentclass[aps,prl,superscriptaddress,
  rsi,amsmath,amssymb,
 reprint,
 floatfix,
]{revtex4-2}
\usepackage{graphicx}
\usepackage{dcolumn}
\usepackage{bm}
\usepackage{comment}
\usepackage{lipsum}
\usepackage{textcomp}
\usepackage{gensymb}
\usepackage{color,colordvi}

\begin{document}

\title{Hyper-Auxeticity and the Volume Phase Transition of Polymer Gels}
\date{\today}

\author{Andrea Ninarello}
\email{andrea.ninarello@cnr.it}
\affiliation{CNR Institute of Complex Systems, Uos Sapienza, Piazzale Aldo Moro 2, 00185, Roma, Italy}
\affiliation{Department of Physics, Sapienza University of Rome, Piazzale Aldo Moro 2, 00185 Roma, Italy}
\author{Emanuela Zaccarelli}
\email{emanuela.zaccarelli@cnr.it}
\affiliation{CNR Institute of Complex Systems, Uos Sapienza, Piazzale Aldo Moro 2, 00185, Roma, Italy}
\affiliation{Department of Physics, Sapienza University of Rome, Piazzale Aldo Moro 2, 00185 Roma, Italy}

\begin{abstract}
Thermoresponsive hydrogels exhibit reversible deswelling at the Volume Phase Transition (VPT), associated to a minimum of the Poisson's ratio $\nu$. Recent numerical investigations uncovered the occurrence of a Hyper-Auxetic Transition (HAT) ($\nu=-1$) for low-crosslinked hydrogels at low temperature, accompanied by a critical-like behavior. Here, we perform extensive numerical simulations to unveil the relation between these two transitions. 
We find that the HAT occurs at different temperatures $T$ up to a maximum value of the crosslinker concentration $c$, thus being clearly distinct from the VPT, taking place at fixed $T$ and for all $c$. Our results provide novel fundamental insights on the interplay between network collapse and mechanical properties of these fascinating materials.
\end{abstract}

\maketitle

Chemical hydrogels are covalently crosslinked polymer chains dispersed in a water-based solvent, offering versatility in synthesis and functionality, driving both scientific inquiry and practical applications~\cite{Hydrogels1, Hydrogels2}. Thermoresponsive hydrogels, like those based on Poly(N-isopropylacrylamide) (pNIPAM), have gathered significant attention for their reversible swelling behavior induced by a change of temperature $T$. This is due to a variation of the affinity between polymer and solvent, usually water, which decreases, becoming more hydrophobic, upon increasing $T$~\cite{Koetting2015}. Experimental studies on neutral and charged pNIPAM hydrogels have elucidated this phenomenon, revealing a Volume Phase Transition (VPT) and its theoretical underpinnings within the Flory-Rehner theory of swelling~\cite{Onuki1993, Dimitriyev2019}. 
\\
Mechanical instabilities are found to accompany the transition~\cite{Tanaka1987,Matsuo1992, Chang2018}, showcasing intriguing elastic properties and the onset of auxetic response, where materials expand perpendicularly with respect to the applied strain, yielding a negative Poisson's ratio that is a signature of auxetic behavior~\cite{Hirotsu1991, Li1993, Boon2017}. Auxetic materials often captivated the interest of researchers for their theoretical implications and practical applications, arising either from material topology or from the proximity to a thermodynamic transition ~\cite{Lakes1987, Evans1991, Xinchun2007, Greaves2011, Nicolaou2012, Ciarletta2013, Pigowski2017, Reid2018, Hanifpour2018}.

In general, when approaching the lowest limit of mechanically stable unconstrained solids ($\nu = -1$), a condition that we name as hyper-auxeticity, the bulk modulus tends to zero, suggesting the presence of very large volume fluctuations from a thermodynamic viewpoint~\cite{Greaves2011}. However, according to the mean field theory of polymer networks, the relationship $K+4/3G=0$ is expected to hold at the VPT~\cite{Tanaka1977,Tanaka1978}, implying a positive $\nu$ within linear elasticity theory, as $\nu=(3K-2G)/[2(3K+G)]$. This prediction is however inconsistent with experimental observations. While simulations could aid in reconciling this disparity, so far most computational studies concerning the VPT of hydrogels have overlooked the interactions between polymers and solvents and the link between thermodynamic and mechanical response~\cite{Escobedo1996, Escobedo1997, Escobedo1997b, Jha2011}.

Recent numerical works~\cite{Ninarello2022,Ninarello2023} have uncovered hyper-auxetic behavior of polymer networks in good solvent, hence purely repulsive systems. This condition is met under a slight tension and by reduction of the crosslinker concentration $c$, leading to the occurrence of 
 a thermodynamic criticality, that is  referred as Hyper-Auxetic Transition (HAT)~\cite{Ninarello2023}. However, in that case, the role played by the attraction 
 and the influence of the deswelling phenomenon on the elastic properties remained unexplored.

In the present work, we fill this gap by performing extensive computer simulations to investigate the occurrence of the HAT in polymer networks experiencing temperature-induced shrinking in a wide range of phase diagram and for several values of the crosslinker concentration. To this aim, we rely on a widespread bead-spring model of polymer networks~\cite{Grest1986,Grest1990} and we  model implicitly the presence of the solvent via a solvophobic interaction~\cite{Soddemann2001}, that was previously shown to quantitatively reproduce the features of the networks across the VPT~\cite{Ninarello2019}.
 
Our findings reveal the presence of an attraction-induced HAT consistent with the Ising universality class, which vanishes with increasing $c$. In addition, we are able to clearly distinguish the solvent-mediated network collapse driven by the VPT and the hyper-auxetic criticality, which at low enough $c$ takes place in different regions of the phase diagram. 
This is in contrast to the VPT, which always occurs at a fixed temperature, that is dictated by the underlying monomer-solvent affinity. Our study thus establishes a clear relationship between thermodynamic and mechanical properties of polymer networks, paving the way to their fine control in experiments.

We simulate hydrogels using the well-established Kremer-Grest model for monomer-monomer interactions, where beads of diameter $\sigma$ experience a Weeks-Chandler-Andersen (WCA) soft repulsion and are linked by finitely extensible nonlinear elastic (FENE) springs~\cite{Grest1986,Grest1990}. The network comprises polymer strands connected by four-valence crosslinkers. The majority of initial configurations are prepared by adding monomers at the equilibrium bond length in a diamond-like crystalline structure (ordered systems), while for one case we also perform self-assembly of patchy particles (disordered systems). This latter method has been recently envisaged to create low-density disordered polymer networks~\cite{Gnan2017,Sorichetti2021,Sorichetti2023}.
We implicitly account for the solvent by introducing an attractive term between all polymer beads in the network, known as the solvophobic potential~\cite{Soddemann2001}, where the strength of the monomer-monomer attractive interactions is controlled by a parameter $\alpha$. We anticipate our findings to align with those of explicit solvent simulations, as demonstrated previously with microgel particles~\cite{Camerin2018}. 

We perform NPT molecular dynamics simulations of the hydrogels at different $c$, $P$ and $\alpha$ values using the LAMMPS package~\cite{Plimpton1995} either in unperturbed conditions or under uniaxial strain. We then calculate the bulk modulus directly from volume fluctuations as $K=\frac{k_bT<V>}{<V^2>-<V>^2}$, while the Young modulus and the Poisson's ratio are computed though their definitions in strain-stress simulations ($Y=\frac{\sigma_\parallel}{\lambda_\parallel}$, $\nu=-\frac{\lambda_\perp}{\lambda_\parallel}$), where $\sigma_{\parallel,\perp}$ and $\lambda_{\parallel,\perp}$ are respectively the stress and the strain either parallel or perpendicular with respect to the deformation axis. Strain-stress simulations consist in uniaxial deformations of different amounts, followed by equilibrium simulations spanning $10^6$ steps. 
We perform deformations in each of the three spatial dimensions in separate runs and then average the results. The time unit is deﬁned as $\tau = \sqrt {m\sigma^2/\epsilon}$, where $m$ is the monomer mass and $\epsilon$ sets the energy scale. In addition to the hydrogels, we also perform NVT simulations of chains of various lengths interacting with the same potential, in order to unveil the role of the VPT. Further details about the simulations  can be found in the SI.

\begin{figure}[th!]
\centering
\begin{minipage}[b]{\linewidth}
\includegraphics[width=0.98\textwidth]{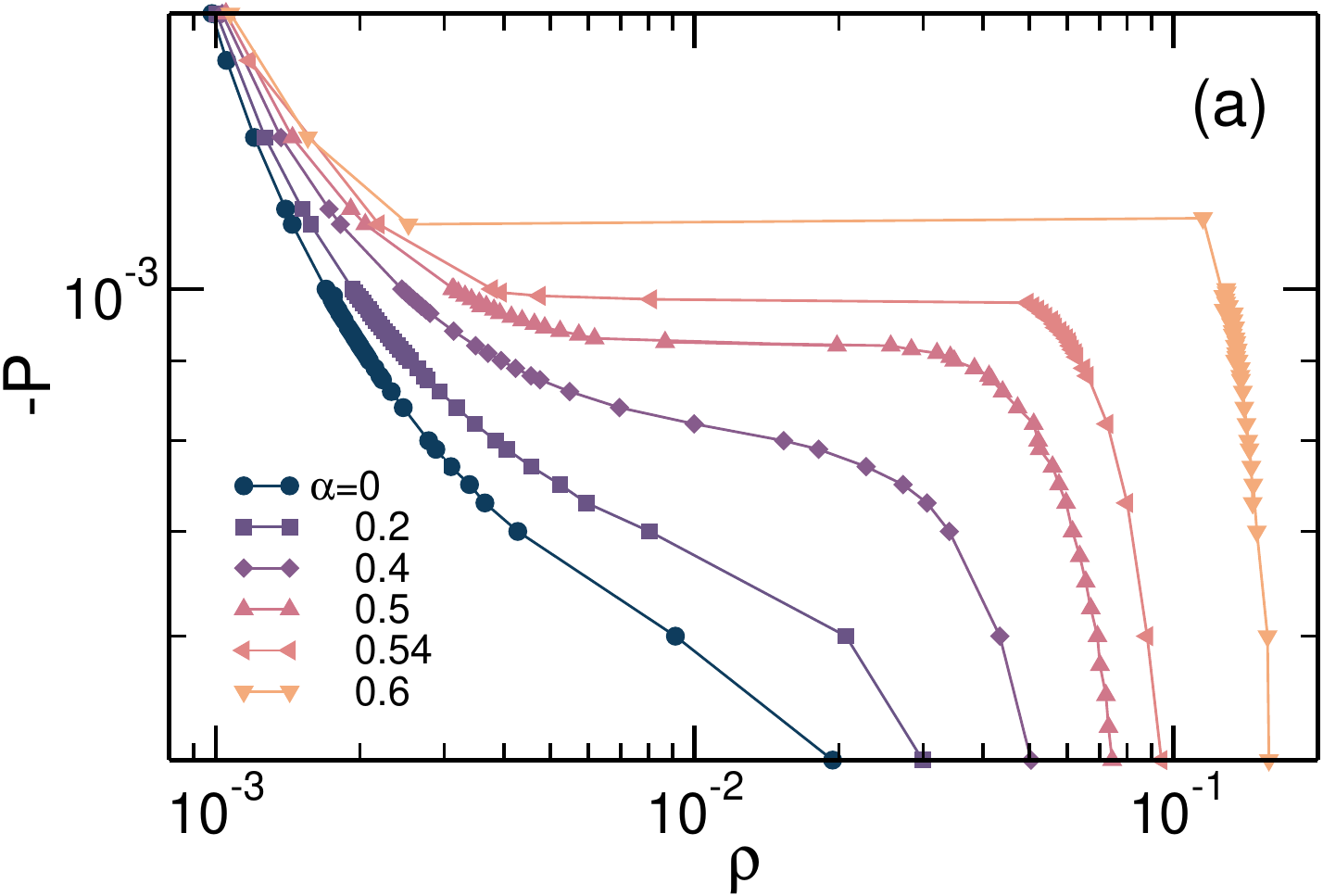}
\hspace{0.02\textwidth}
\end{minipage}
\begin{minipage}[b]{0.48\linewidth}
\includegraphics[width=1.\textwidth]{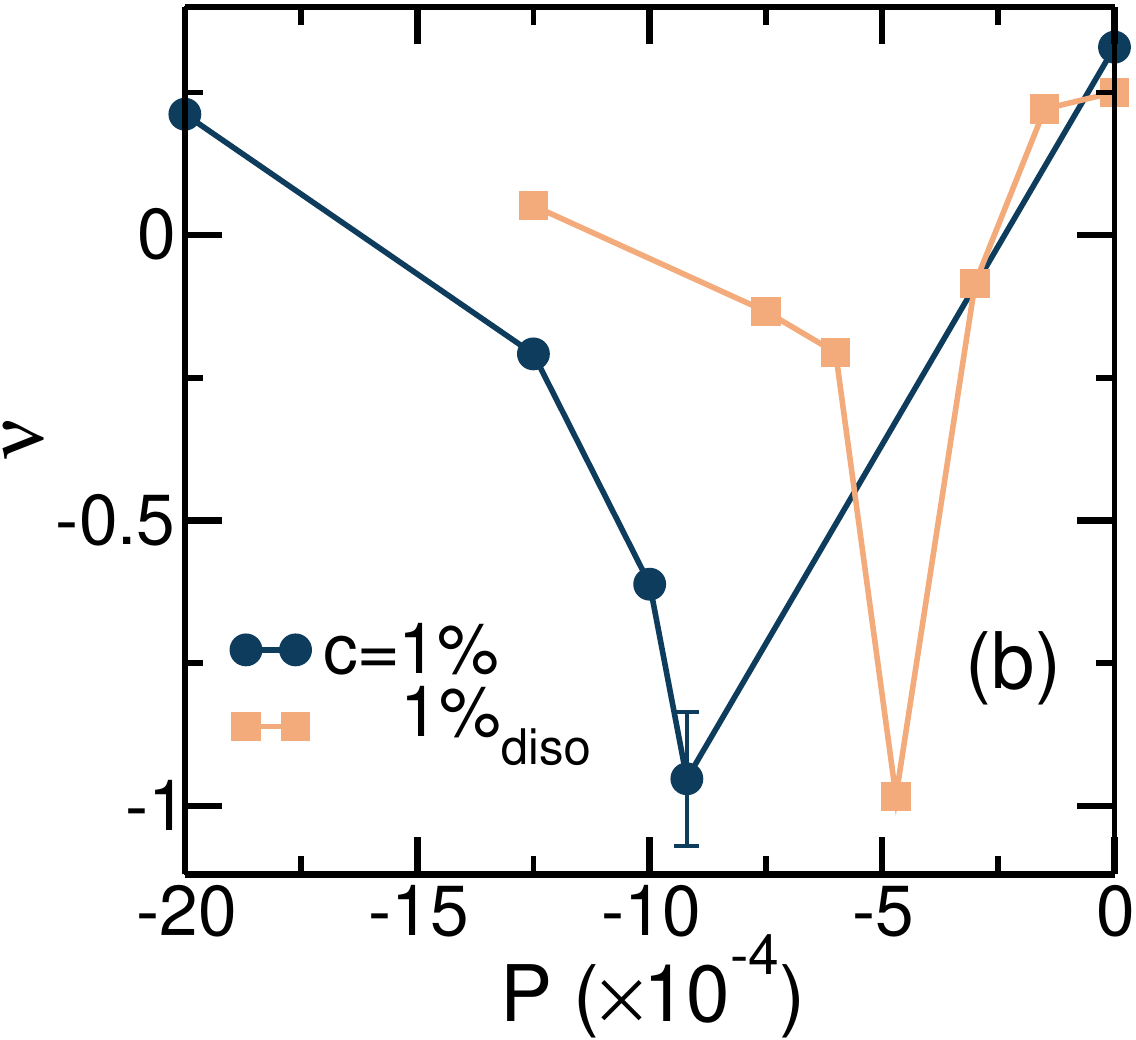}
\end{minipage}
\begin{minipage}[b]{0.47\linewidth}
\includegraphics[width=1.\textwidth]{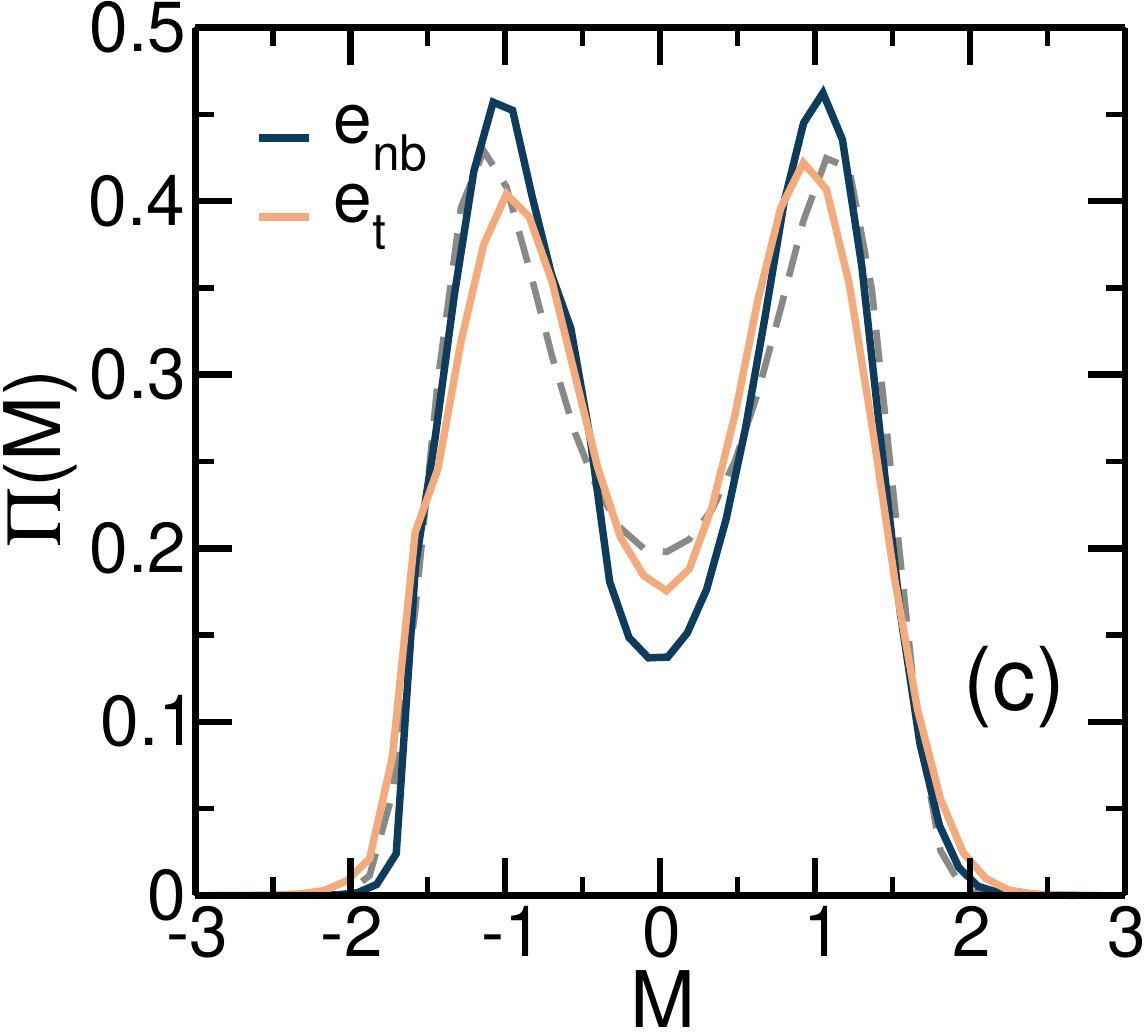}
\end{minipage}
\\
\caption{Hyper-auxeticity in solvent-mediated networks: (a) Equation of state representing the negative pressure $P$ as a function of $\rho$ for different attraction strengths $\alpha$ as indicated in the legend for the $c=1\%$ system. (b) Poisson's ratio $\nu$ as a function of the pressure for the $c=1\%$ (blue circles) and $1\%_{\text{diso}}$ (orange squares) systems respectively at $\alpha=0.5$ and $0.425$. (c) Normalized probability distribution $\Pi(M)$ of the order parameter $M=\rho+s e_X$, using either total energy $e_T$ (blue line) or non-bonded particles energy $e_{nb}$ (orange line). Curves are rescaled to zero-mean and unitary variance. Results are for the $c=1\%$ diamond network at $\alpha_{HAT}=0.5064$, reweighted to $P=9.319\times10^{-4}$, and using the mixing parameter $s=-3.2, -0.8$ for the two energies, respectively.}
\label{fig:aux_c1}
\end{figure}

We start by investigating the onset of auxetic behavior in the presence of attraction. For this purpose, we analyze the equation of state of the ordered system with $1\%$ crosslinking, varying tension (or negative pressure $-P$) for different temperatures (quantified by the attraction strength $\alpha$). The corresponding findings are reported in Fig.~\ref{fig:aux_c1}(a). Notably, as the effective attraction between particles increases, the change in the  density of the network becomes more abrupt, leading to a discontinuous transition between two states, characterized by a low and a high density, respectively, when $\alpha > 0.5$. Given the resemblance to a classical liquid-gas transition, we  examine in more detail the system behavior at $\alpha=0.5$ for different pressures. \\
Our analysis of the elastic moduli reveals a non-monotonic behavior of the bulk $K$ and Young $Y$ moduli as pressure changes (see SI) that reflects the behavior of the Poisson's ratio $\nu$, reported in Fig.\ref{fig:aux_c1}(b). First, $\nu$ decreases with decreasing $P$, reaching negative values, and then it increases again at larger tensions. Notably, these two behaviors are distinguished by a minimum where $\nu=-1$, marking the mechanical stability limit~\cite{Milton1992, Lakes1993}. We identify this condition as a Hyper-Auxetic point, that in addition to the mechanical instability, is characterized by huge density fluctuations, reminiscent of  critical-like behavior. To quantify this behavior in the hypothesis of the occurrence of a second-order critical point, we follow the typical approach used to study gas-liquid phase separation in atomic and molecular systems~\cite{Wilding1992, Debenedetti2020}. We thus monitor the behavior of the characteristic order parameter $M$, which combines density and energy of the system, akin to approaches used in liquid-gas transitions~\cite{Wilding1992}. Hence, $M=\rho+se_X$, where $s$ is a mixing parameter and $X$ identifies the type of energy employed in the analysis. In particular, we either employ the total energy $e_t$ or the energy of the non-bonded particles $e_{nb}$ only, as in our previous studies~\cite{Ninarello2022,Ninarello2023}. By examining the probability distribution of this order parameter rescaled to zero-mean and unitary variance $\Pi(M)$, we compare it to the theoretical Ising behavior and find a good agreement for the critical temperature $\alpha_{HAT}^{1\%}=0.5064$, between the predicted and calculated histograms with both observables, as shown in Fig.\ref{fig:aux_c1}(c). We ascribe the differences between the results obtained with either $e_t$ or  $e_{nb}$ to the intrinsic difficulties in finding these distributions in computational studies and to the weak correlation between energy and density for  the considered low-density polymer network. Our results thus suggest that the HAT transition observed in the presence of attraction is of the Ising universality class, deepening the analogy to a gas-liquid phase separation.

To provide robustness to our findings, we perform a similar analysis also for the disordered $c=1\%$ network ($1\%_{\text{diso}}$). We thus evaluate the equation of state (not shown), establishing the critical-like temperature $\alpha_{HAT}^{1\%_{\text{diso}}}=0.425$, that is notably distinct from the corresponding diamond network one. We then compute the elastic properties, confirming the occurrence of the HAT where $\nu=-1$ at an intermediate pressure, as also shown in Fig.\ref{fig:aux_c1}(b).
Similarly to our previous studies at $\alpha=0$~\cite{Ninarello2022,Ninarello2023}, we thus confirm that the topology of the network does not play a role for the existence of the HAT, ruling out a possible geometric origin.

\begin{figure}[t]
\centering
\begin{minipage}[b]{\linewidth}
\includegraphics[width=0.95\textwidth]{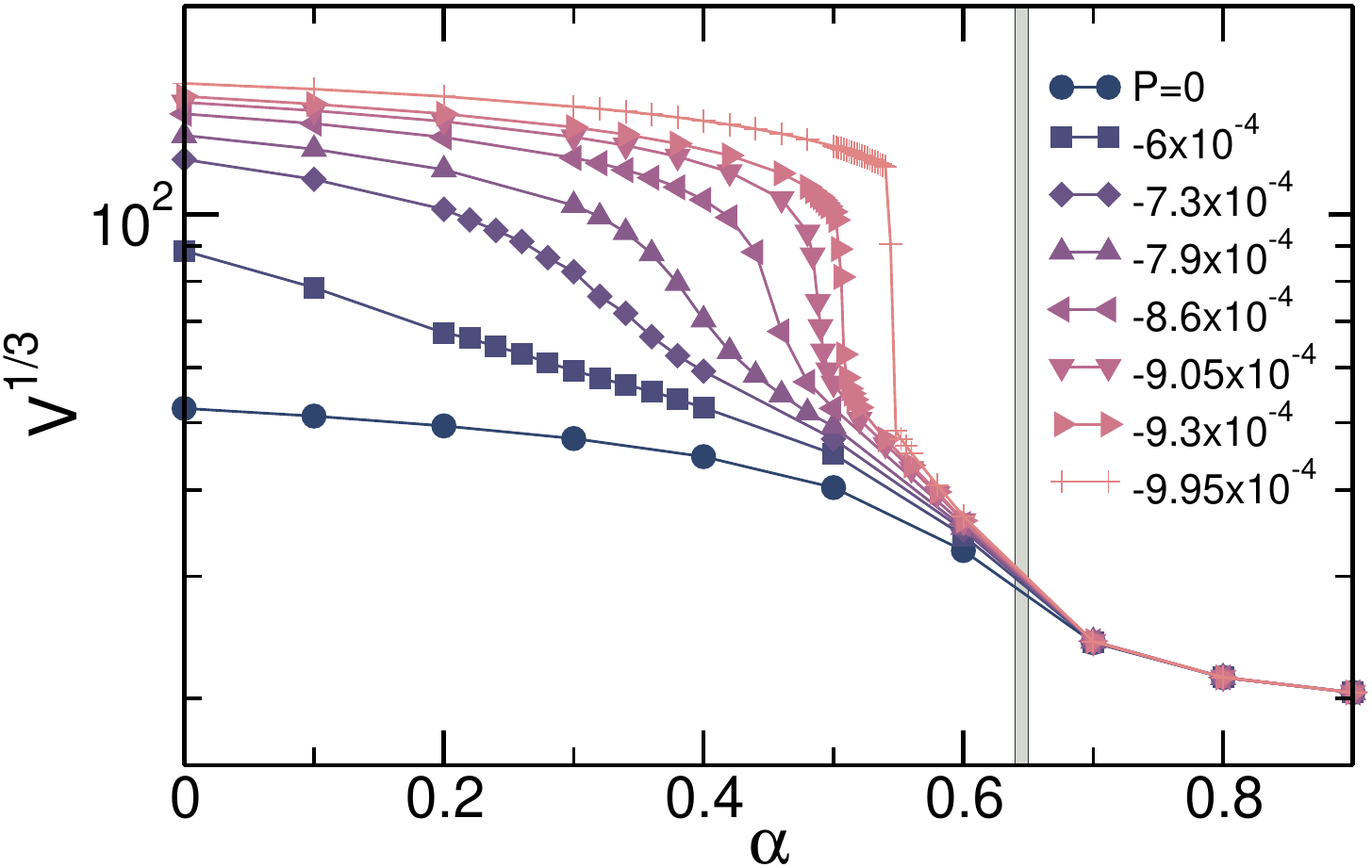}
\end{minipage}
\begin{minipage}[b]{\linewidth}
\includegraphics[width=0.444\textwidth]{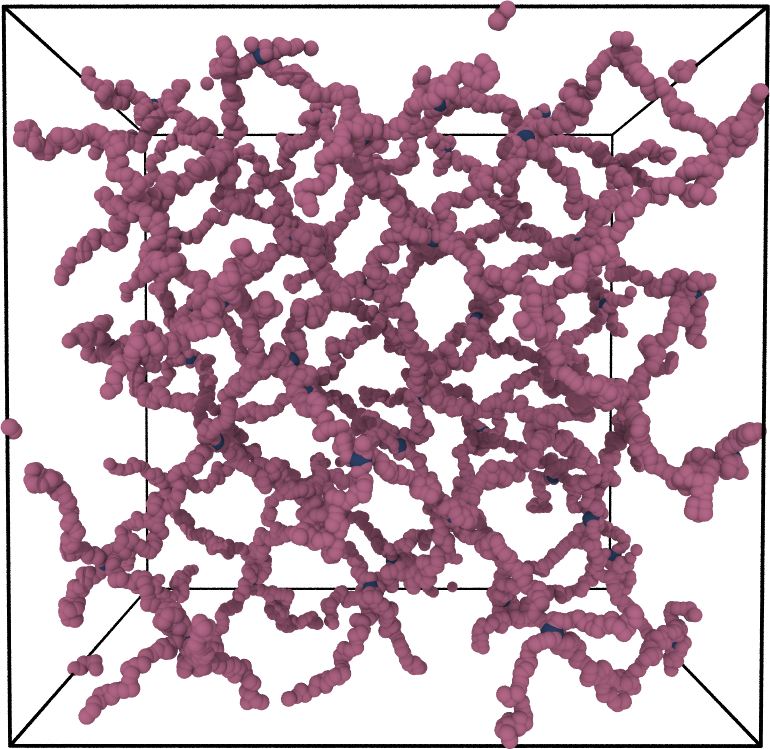}
\includegraphics[width=0.47\textwidth]{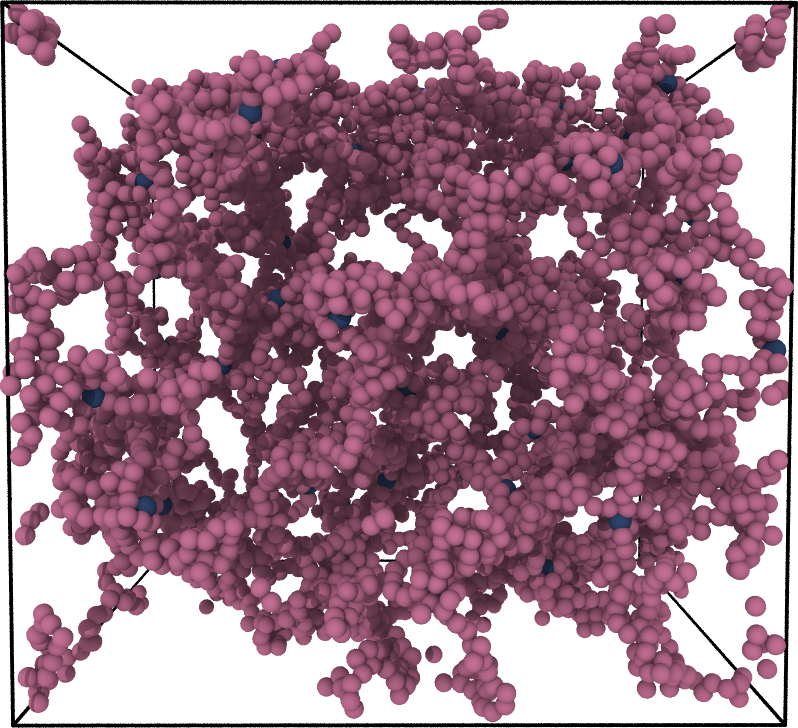}
\end{minipage}
\caption{Top: Swelling curve for the $c=1\%$ ordered network. Cubic root of the volume as a function of $\alpha$ for different $P$ as indicated in the legend. The vertical region indicates the location of the VPT. Bottom: Snapshots of the $c=1\%$ system at $P=-9.05\times10^{-4}$ around the HAT, for $\alpha=0.48$ (left) and $0.56$ (right), respectively. Monomers and crosslinkers are shown in violet and dark blue, respectively.}
\label{fig:swelling}
\end{figure}

We now aim to compare these results with  previous measurements on hydrogels, reporting the occurrence of auxetic behavior~\cite{Hirotsu1991}. Indeed, at first sight, it may appear that the Hyper-Auxetic Transition (HAT) is merely an amplification of the negative Poisson's ratio observed in experiments~\cite{Hirotsu1991}. However, a more quantitative analysis is needed to verify this assumption, specifically examining the relationship between the two transitions: the HAT and the VPT.
To address this issue, we report the hydrogel swelling curve for the $c=1\%$ diamond network in Fig.~\ref{fig:swelling}, where the cubic root of volume $V^{1/3}$, analogous to an effective size of the hydrogel, is plotted as a function of $\alpha$, which acts as an effective temperature, for different studied pressures. First, we focus on $P=0$, that is the case representative of microgels in experiments, and observe that, as usual, the effective size  smoothly decreases with increasing $\alpha$, showing an inflection point at the VPT, namely at $\alpha_{VPT}\simeq 0.65$~\cite{Gnan2017}. However, upon decreasing $P$, we find that deswelling happens in two steps. At first, there is a decrease which progressively hits the HAT point at $\alpha_{HAT} \sim 0.5$, above which $V^{1/3}$ becomes discontinuous. Representative snapshots of the hydrogel around the HAT are also reported in Fig.~\ref{fig:swelling}, showing the spinodal-like character of the transition. Indeed, the high density state is far from homogeneous and a further compaction only happens at larger $\alpha$. This second deswelling step finally recovers the VPT behavior at high temperatures, where for all pressures the deswelling curves are virtually indistinguishable from one another.

These results clearly indicate that the HAT driving the auxetic response of the system is a distinct phenomenon from the VPT and we now aim to unveil their underlying nature and mutual interplay in more detail.

\begin{figure}
\centering
\includegraphics[width=0.48\textwidth]{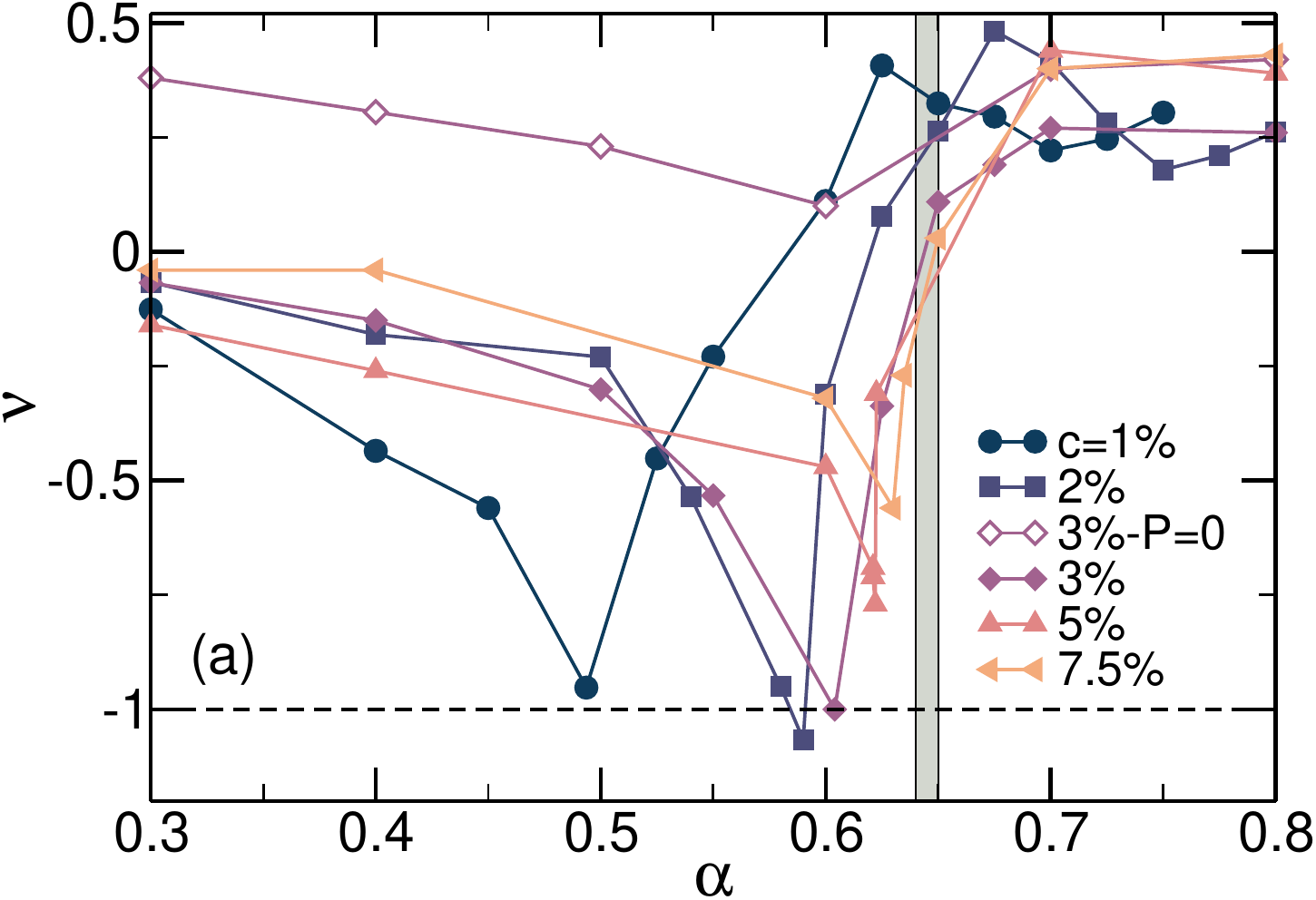} 

\vspace{0.1cm}

\includegraphics[width=0.455\textwidth]{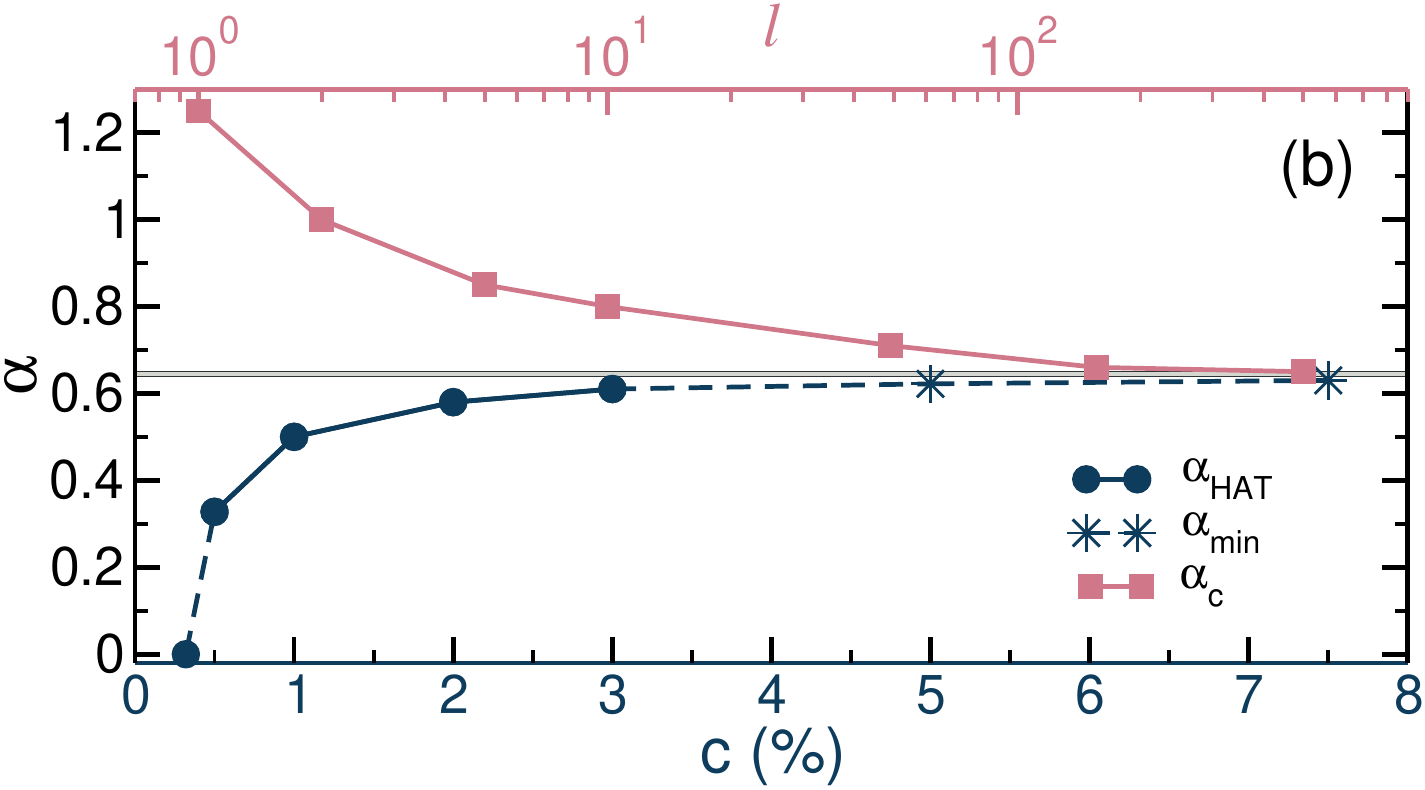}
\caption{HAT and VPT. (a) Poisson's ratio as a function of $\alpha$ for ordered systems with $c=1,2,3,5,7.5\%$ and corresponding $P= 9.25\times10^{-4}, 3.5\times10^{-3}, 7.9\times10^{-3},1.8\times10^{-2},4.25\times10^{-2}$. The horizontal dashed black line indicates the hyper-auxetic threshold, when $\nu_{min}=-1$, while the vertical box marks $\alpha_c$, i.e., the occurrence of the VPT.  (b) Top. The phase separation effective temperature of chains $\alpha_c$ is reported as a function of chain length $l$. Bottom. Effective temperature, corresponding to the minimum in $\nu$ as a function of $c$. Filled circles indicate HAT points where $\nu=-1$, while stars represent points where $\nu \gtrsim -1$. Both top and bottom curves tend to the VPT one, $\alpha_{VPT}$, as $c$ and $l$ increase, as signaled by the horizontal line.}
\label{fig:HAT_to_VPT}
\end{figure}

To this aim, we first examine the occurrence of the HAT for various crosslinker concentrations. We explore diamond networks with $c=0.5,1,2,3,5,$ and $7.5\%$ and investigate their elastic behavior in order to identify the conditions corresponding to the minimum Poisson's ratio $\nu_{min}$.  We first retrieve the corresponding pressure $P_{min}$, that, as expected, increases in absolute value for stiffer networks. Then, we calculate $\nu$ as a function of $\alpha$ for each network in order to determine whether the system is able to reach $\nu_{min}=-1$ independently of crosslinker concentration. 
The results of this extensive analysis are summarized in Fig.~\ref{fig:HAT_to_VPT}(a), where only some of the investigated $c$ values are reported to improve visualization. 
Our findings reveal the occurrence of hyper-auxeticity at a given $\alpha_{HAT}$, associated to $\nu_{min}\approx -1$, up to crosslinker concentrations $c
\simeq 3\%$, while networks with higher $c$ are still found to exhibit a minimum in $\nu$, at a characteristic temperature that we call $\alpha_{min}$, which however does not reach the HAT condition any longer, since $\nu_{min}>-1$. Indeed, in analogy with previous results obtained for $\alpha=0$~\cite{Ninarello2022} and for microgels~\cite{Rovigatti2019}, we find that $\nu_{min}$ increases with the stiffness of the network, approaching the HAT condition only at sufficiently low $c$. 
It is now interesting to look at the $\alpha_{HAT}$ and $\alpha_{min}$ dependence on crosslinker concentration, which is reported in Fig.~\ref{fig:HAT_to_VPT}(b). It is evident that  $\alpha_{HAT}$ continuously turn to $\alpha_{min}$, progressively increasing with the system connectivity and finally tending to $\alpha_{VPT}\sim 0.65$ at high $c$.
Hence, while the HAT terminates at a finite $c$ value, we provide evidence that its echo, characterized by the presence of a minimum in $\nu$ at even larger $c$, converges to the temperature of the VPT as crosslinker concentration increases. This is why a minimum in $\nu$ is detected even in microgels~\cite{Voudouris2013, Boon2017, Rovigatti2019}, a feature here confirmed for hydrogels at $P=0$, as also shown for the $c=3\%$ network in Fig.~\ref{fig:HAT_to_VPT}(a).
\\
Interestingly, we previously detected the existence of the HAT at $\alpha=0$ in the $c=0.35\%$ network~\cite{Ninarello2022} in the absence of attraction. Although in that case we found comparable phenomenology with the present investigation, here we notice that the two phenomena appear to be continuous from the behavior of $\alpha_{HAT}$ in Fig.~\ref{fig:HAT_to_VPT}(b). However, the underlying order parameter in that case was found to be strictly $\rho+e_{nb}$, so that differences between the attractive and purely repulsive case need to be further investigated.
\\
Having discussed the HAT, we now address the underlying nature of the VPT. We thus consider simple polymer chains featuring the same interactions as the hydrogels and evaluate the occurrence of gas (chain poor)-liquid (chain rich) phase separation at a given critical temperature $\alpha_c$, as detailed in the SI, for systems of different chain lengths $l$ from monomers ($l=1$)  to very long chains ($l\sim 500$). We find that, as expected, the various systems phase separate at a temperature which depends on $l$. In particular, a larger temperature is needed for monomers and small chains, while  connectivity enhances the tendency to phase separate, thus lowering the critical temperature as $l$ increases. Remarkably,  for long enough chains, i.e. $N\gtrsim 100$, we find that phase separation occurs at a constant value $\alpha_c \sim \alpha_{VPT}$. Given the presence of very long chains in any microgel or hydrogel realization, either in experiments or in simulations, this finding explains the almost universal value of $\alpha_{VPT}$ and of the corresponding temperature in experiments, that occurs independently of $c$ for PNIPAM-based systems in the absence of co-monomers. This is simply due to the solvophobic interactions between polymer and water, which drives the phase separation and is mitigated by the presence of crosslinks within the network, which transforms the critical-like behavior into a smooth and reversible deswelling process. 

In summary, the present study elucidates the multifaceted relationship between critical swelling, hyper-auxeticity, and Volume Phase Transition in thermoresponsive polymer networks. Through extensive numerical  simulations in equilibrium and under uniaxial deformation, we demonstrate that the Hyper-Auxetic Transition represents a clearly distinct phenomenon from the $VPT$, that is essentially driven by the phase separation of polymer chains, echoing the coil-to-globule transition of single chains. By analyzing the swelling curves of the hydrogels, we are indeed able to discern a distinct two-step process at slightly negative pressures, whereby an initial compression leads to the onset of a hyper-auxetic point before a further volume decrease takes place, toward the generic $VPT$ regime. Furthermore, our investigation spanning a wide range of crosslinker concentrations uncovers hyper-auxetic behavior up to a finite $c\simeq 3\%$, while higher crosslinker concentrations exhibit increasing stiffness without reaching the mechanical instability. The convergence of the Hyper-Auxetic Transition with the $VPT$ as system connectivity increases underscores the complex nature of polymer network behavior. Future theoretical work will have to be devoted to the understanding of the link between the hyper-auxetic transition found in this work in presence of attractive (solvophobic) interactions and the one found at $\alpha=0$~\cite{Ninarello2022}, where only entropy plays a role, in order to shed light on the fundamental nature of this transition from the statistical mechanics point of view. 
Notwithstanding this open theoretical problem, the present study provides a useful framework for the understanding of previous and future experimental observations thanks to the establishment of a clear link between thermodynamic and mechanical properties of polymer networks. 
By clarifying the conditions under which the HAT occurs, the present results call for further experimental and computational exploration of elastic and thermodynamic properties of polymer networks, both at negative and at positive pressures, paving the way for the development of novel materials with controlled auxetic properties.

 We acknowledge funding from ICSC – Centro Nazionale di Ricerca in High Performance Computing, Big Data and Quantum Computing, funded by European Union – NextGenerationEU - PNRR, Missione 4 Componente 2 Investimento 1.4 and CINECA-ISCRA for computational resources.

\bibliography{submit}
\bibliographystyle{apsrev4-2}

\newpage
\onecolumngrid

\section*{Supplementary Information for: Hyper-Auxeticity and the Volume Phase Transition of Polymer Gels}

\subsection*{Additional Methods}
We perform Molecular Dynamics simulations of polymer networks composed of monomers that interact through the Kremer-Grest potential. Excluded volume for all particles are given by the Weeks-Chandler-Andersen potential:~\cite{Weeks1971}
\begin{equation}
\label{eq:WCA}
V_{WCA}\left( r \right )= 4\epsilon\left[\left(\frac{\sigma}{r} \right )^{12}-\left(\frac{\sigma}{r} \right )^{6} \right ]+\epsilon \ \ \  if \ \ \  r \leq 2^{1/6}\sigma,\\
\end{equation}
where $\sigma$ is the monomer diameter, which sets the unit of length, and $\epsilon$ controls the energy scale. Defining $m$ as the mass of the particles, the unit time of our simulations is defined as $\tau=\sqrt{m\sigma^{2}/\epsilon}$. Chemical bonds between connected monomers are modeled by a FENE potential $V_{FENE}\left(r\right)$~\cite{Kremer1990}:

\begin{equation}
\label{eq:FENE}
V_{FENE}\left( r \right )=-\epsilon k_{F}R^{2}_{0}\ln\left[1-\left(\frac{r}{R_{0}\sigma} \right ) \right ] \ \ \  if  \ \ \  r < R_{0}\sigma,\\
\end{equation}

where $k_{F}=15$ is the spring constant and $R_{0}=1.5$ is the maximum extension of the bond.
To simulate the presence of the solvent and the polymer progressively going from a  hydrophilic to a hydrophobic condition with increasing temperature, we employ a so-called solvophobic potential~\cite{Soddemann2001} of the form:
\begin{equation}
  \label{va_force}
  V_\alpha(r)=
  \begin{cases}
    -\epsilon \alpha \ \        &\ r\leq 2^{\frac{1}{6}}\sigma,\\
    \frac{1}{2}\alpha \epsilon  \big[ \cos(\gamma(\frac{r}{\sigma})^2+\beta)-1 \big]  \ \     &\ 2^{\frac{1}{6}}\sigma<r<R_0\sigma,
  \end{cases} 
\end{equation}
where $\gamma=(\pi(2.25-2^{1/3}))^{-1}$, $\beta = 2\pi-2.25\gamma$, $\epsilon$ is the unit of energy and the parameter $\alpha$ controls the strength of the monomer-monomer attractive interactions.

As anticipated in the main text, we employ networks with two different topologies. In one case, we build networks with an ordered structure based on a diamond-like lattice by crosslinking equal length chains where monomers are placed at the equilibrium distance of the FENE potential. The crosslinkers concentration is therefore directly determined by the chain length $l$ through the relationship: $c=1/(2l+1)$. 
The total number of monomers $N$ of the networks is $N=\frac{N_c}{c}$ where $N_c$ is the number of crosslinkers. In particular, we use $N_c= 64$ and $l=99,49,25$, respectively for the $c=0.5\%,1\%,2\%$ systems, while $N_c= 216$ and $l=16,10$ for the $c=3\%,5\%$ systems, and finally $N_c=512$ and $l=6$ for the $c=7.5\%$ system.

In the disordered case, we considered a network with $N=4384$ total number of monomers, of which $N_c=49$ crosslinkers. This is assembled through a recently developed technique~\cite{Gnan2017}, which makes us able to produce low density computational polymer networks. In this method, two and four valence patchy particles that mimic respectively monomers and crosslinkers are mixed in appropriate proportions in a cubic box with periodic boundary conditions to obtain given connectivity amount. The assembly is performed via NVT molecular dynamics simulations at low temperature $T=0.3$ and density $\rho=2.85\times10^{-2}$ with the OxDNA package~\cite{Rovigatti2014, Poppleton2020}. We wait until the $99.9\%$ of possible bonds are satisfied, then remove all particles non belonging to the biggest cluster and all dangling chains in order to have a fully connected network without defects. Once the assembly is ready, we then  substitute the patchy reversible bonds with the bead-spring model.

\subsection*{Bulk and Young moduli}

\begin{figure}[h]
\centering
\includegraphics[width=0.48\textwidth]{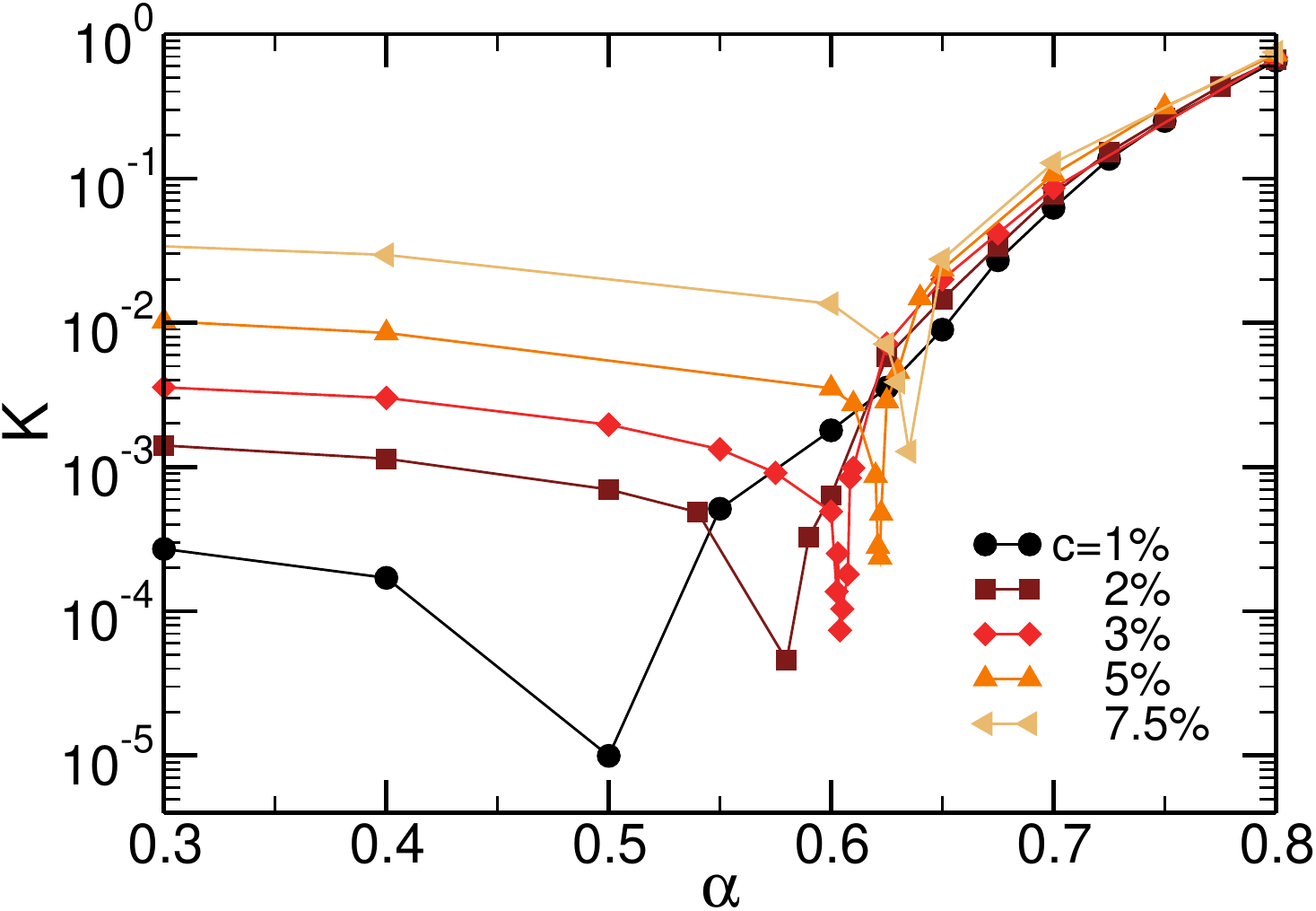}
\includegraphics[width=0.48\textwidth]{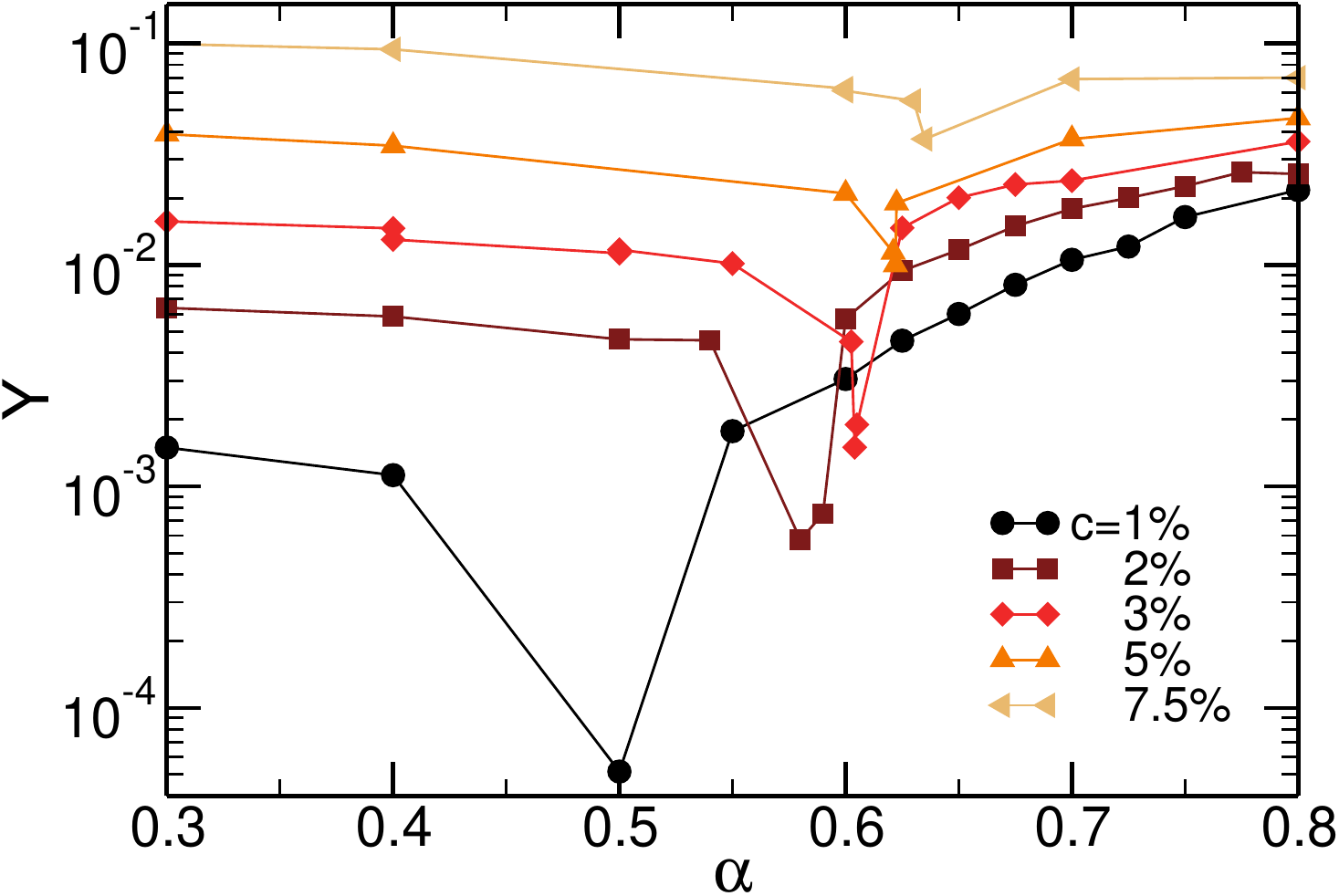}
\caption{Bulk (a) and Young (b) moduli as a function of $\alpha$ for $c=1,2,3,5,7.5\%$ and corresponding $P= 9.25\times10^{-4}, 3.5\times10^{-3}, 7.9\times10^{-3},1.8\times10^{-2},4.25\times10^{-2}$.}
\label{fig:KG}
\end{figure}

As discussed in the main text, the behavior of the bulk $K$ modulus for our hydrogels suggests the presence of an underlying criticality. In particular, a vanishing bulk modulus is related with the presence of critical-like fluctuations of the system volume, hinting at  the occurrence of a second order transition as observed for instance in martensitic transformations~\cite{Dong2010}. 
In Fig.~\ref{fig:KG}(a) we report the bulk modulus as a function of $\alpha$ for different $c$ values, corresponding to state points shown in Fig.3(a) of the main text. From low $\alpha$ results included in Fig.~\ref{fig:KG}(a) , it can be seen that the elasticity of the network highly depends on crosslinker concentration, with values of $K$ spanning two orders of magnitude for increasing $c$. However, increasing $\alpha$, $K$ shows a dramatic decrease with a minimum reaching very low values as $c$ decreases, up to about five orders of magnitude with respect to the high compact state observed at large $\alpha$ when all networks behave in the same way. In addition, we report in Fig.~\ref{fig:KG}(b) the corresponding behavior of the Young modulus $Y$. Despite this is not expected to show critical behavior, we observe the same qualitative trend also present in $K$, with a clear minimum also developing, becoming more and more pronounced as $c$ decreases. Increasing $\alpha$ to values above the VPT, we find that $Y$ increases and remains distinct for different $c$ values, differently from $K$.

\subsection*{Phase separation of polymer chains}
To allow the comparison with the network  results, we also explore chain-based systems of different lengths $l$. We thus perform MD simulations of chains of equal length $l=1,2,5, 10, 49,156, 500$ and varying total number of chains varying in the range  between 10000 and 20000.

To allow a quick scan of the phase diagram in order to look for phase separation, for each chain length we perform several NVT simulations at different attractive strengths $\alpha$ and system volume. After allowing the system to equilibrate for $2\times10^6$ time steps, we generate multiple equilibrium configurations by running simulations for $10^7$ time steps. These configurations are used to calculate the structure factor $S(k)$ of the network, that is computed removing the (almost negligible) contribution of the single chain form factor $S_1(k)$. Subsequently, we extract the value at $S(k)$ at the smallest simulated value of $k$, namely $k_{min}=\frac{2\pi}{L}$, with $L$ representing the box side. In this way, we have access to a proxy for the isothermal compressibility $\kappa_T$ of the system (that is the inverse of the bulk modulus for network systems), since $S(k\to0)$ is directly related to the system through the equation $\kappa_T=\frac{S(0)}{\rho K_BT}$~\cite{Hansen2013}. It is important to note that $\kappa_T$ theoretically approaches infinity at phase separation in infinite-size systems, but for finite systems, density fluctuations remain finite, although showing a significant growth.

\begin{figure}[h]
\centering
\includegraphics[width=0.48\textwidth]{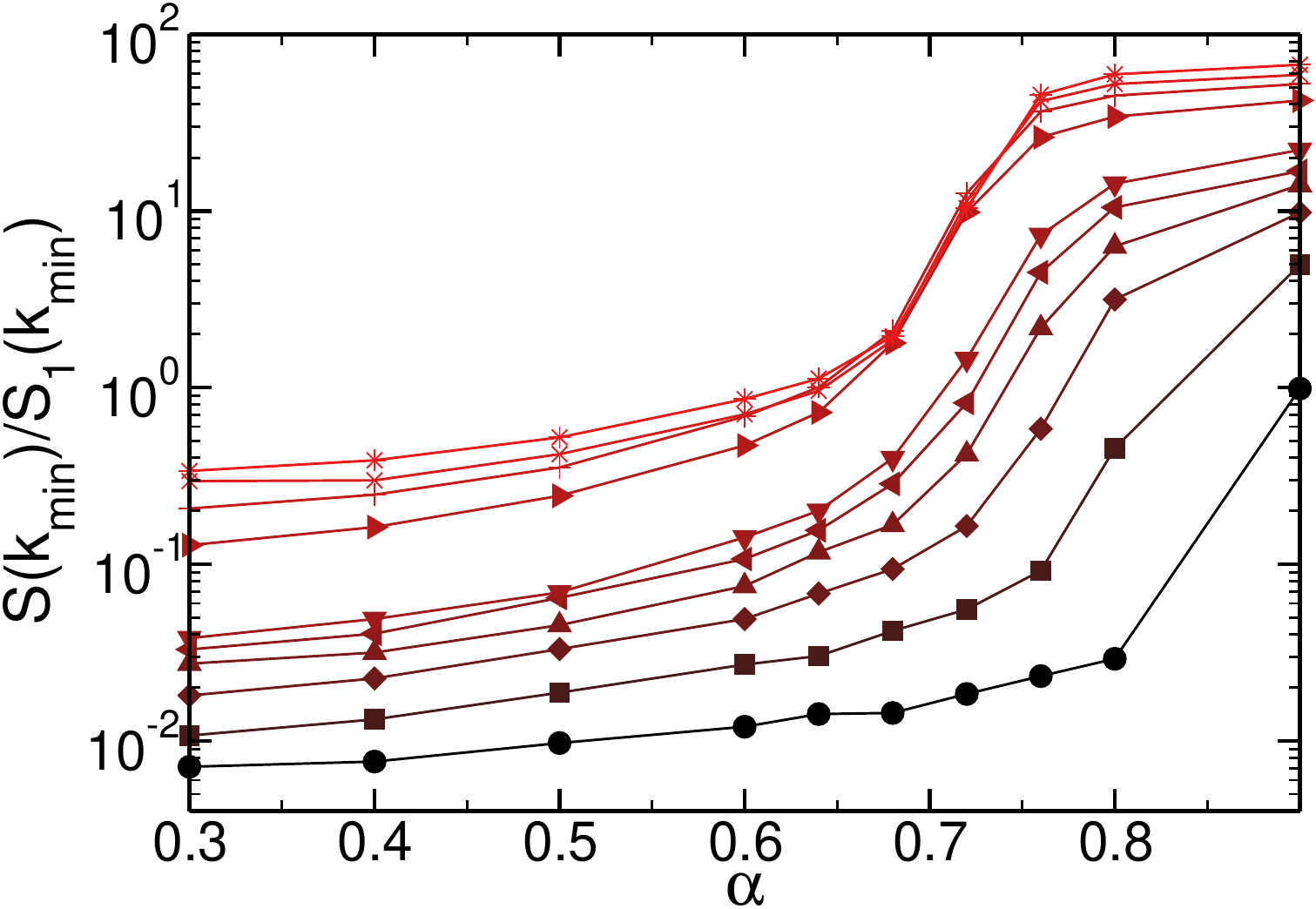}
\includegraphics[width=0.48\textwidth]{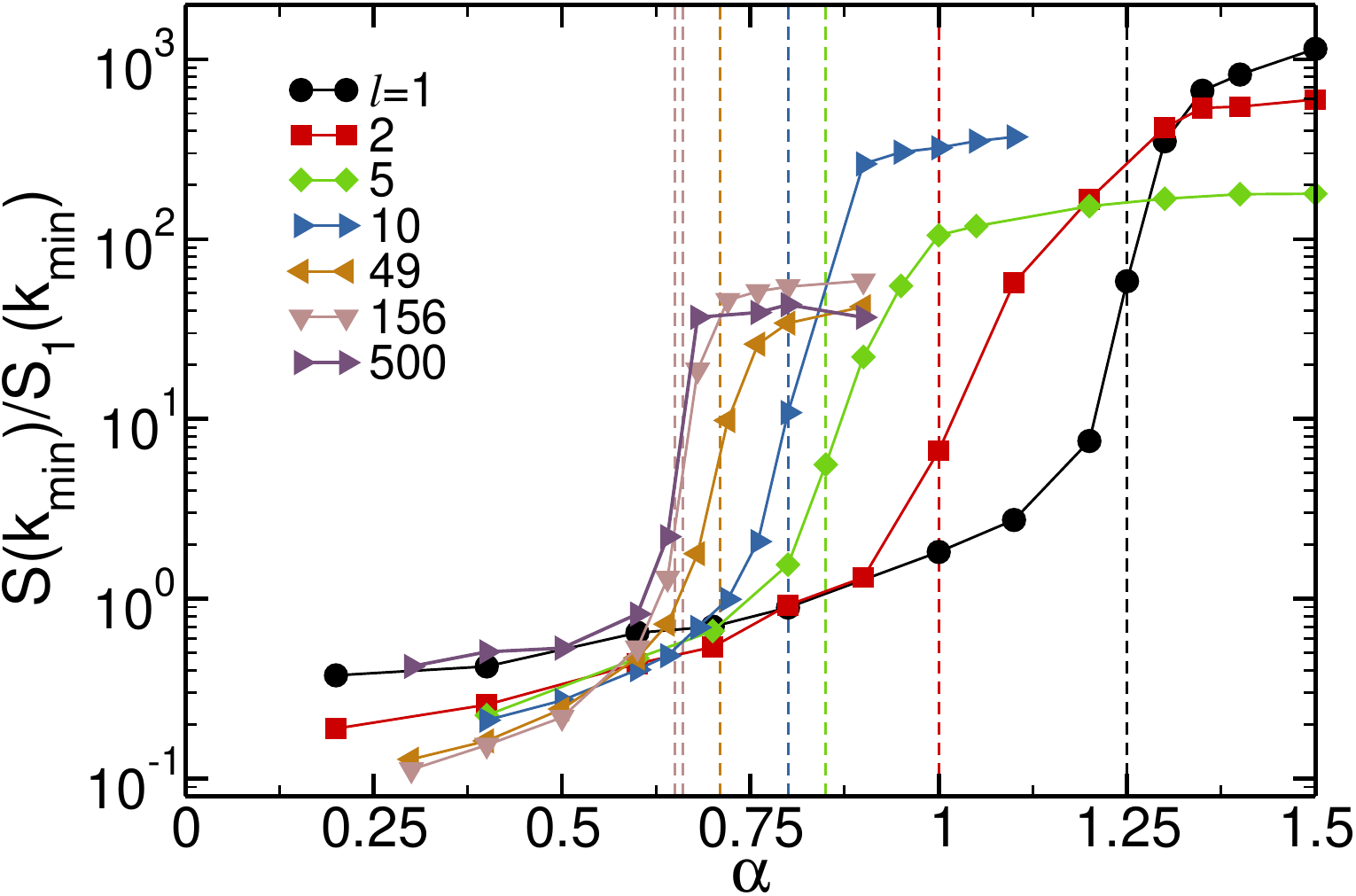}
\caption{Value of the structure factor at the minimum accessible wave vector $S(k_{min})$ as a function of $\alpha$, rescaled by the single chain form factor at the same point $S_1(k_{min})$. (a) Results for the $l=49$ system for different volumes increasing from bottom to top in the interval $V=1\times10^3 - 1\times10^4$. (b) Results for systems with various chain lengths as indicated in the legend at the volume corresponding to the steepest variation. Vertical lines highlight the inflection points that identify the transition temperature $\alpha_c$ for each $l$.
}
\label{fig:skmin_alpha}
\end{figure}

We report $S(k_{min})/S_1(k_{min})$ as a function of $\alpha$ for several studied isochores for the $l=49$ system in Fig.~\ref{fig:skmin_alpha}(a) as a representative example of our analysis.  By increasing $\alpha$, a clear increase of $S(k_{min})$ is observed, which goes through an inflection point at decreasing value of $\alpha$ with increasing volume. At large volumes, the behavior saturates, collapsing onto the same curve, with a value of $S(k_{min})/S_1(k_{min})$ jumping by more than two orders of magnitude for a slight increase of $\alpha$. We identify this inflection point with the critical value $\alpha_c$ that is required to induce phase separation. 
As we aim to compare these results with the network system ones, it is particularly relevant to perform a similar analysis at varying the chain length. This is reported in Fig.~\ref{fig:skmin_alpha}(b) where we report $S(k_{min})/S_1(k_{min})$ for each studied $l$ at the volume where the phase-separation temperature $\alpha_c$ is smallest. We thus
observe a clear decrease of $\alpha_c$ with increasing $l$, approaching the value of the VPT ($\alpha_{VPT}\sim 0.65$) for $l >100$, as shown in the main text in Fig.~3(b). Since networks generally contains mostly longer chains, whose effect is dominating on the phase behavior, we thus conclude that the VPT is an effect purely originating from the solvophobic attraction between the monomers as $\alpha$ increases and its universal value $\alpha_{VPT}$ is due to the dominance of long chains in the systems, both for hydrogels and for microgels.

\end{document}